\documentclass[smallextended]{svjour3}
\pdfoutput=1

\usepackage{algorithmic}
\usepackage{amsmath,amssymb,amsfonts}
\usepackage{cite}
\usepackage{comment}
\usepackage{enumerate}
\usepackage{graphicx}
\usepackage{lipsum}
\usepackage[colorlinks,allcolors=black]{hyperref}
\usepackage{siunitx} 
\usepackage{subfigure}
\usepackage{textcomp}
\usepackage{xcolor}
\usepackage{xspace}
\usepackage{pdflscape}

\newcommand{\RQcause}{RQ1\xspace}
\newcommand{\RQwho}{RQ2\xspace}

\newcommand\PKGGG{\mbox{\scshape\small gender-guesser}\xspace}
\newcolumntype{H}{>{\setbox0=\hbox\bgroup}c<{\egroup}@{}}

\def\dataAuthorsInit{\num{70 778 755}\xspace}
\def\dataAuthorsInitApprox{71\,M\xspace}
\def\dataCommitsInit{\num{3 808 132 978}\xspace}
\def\dataCommitsInitApprox{3.8\,B\xspace}
\def\dataPercUAuthors{61.4\%\xspace} \def\dataPercCommitsByUAuthors{43.7\%\xspace} \def\dataAvgCommitsUAuthors{35.5\xspace} \def\dataAvgCommitsMAuthors{77.5\xspace} \def\dataAvgCommitsFAuthors{53.9\xspace} \def\dataAuthorsPostCleanup{\num{51 664 829}\xspace}
\def\dataAuthorsPostBots{\num{51 617 013}\xspace}
\def\dataAuthorsPostMerge{\num{40 913 389}\xspace}
\def\dataAuthorsPostMergeApprox{41\,M\xspace}
\def\dataAuthorsGenderM{\num{13 354 546}\xspace} \def\dataAuthorsGenderW{\num{2 398 896}\xspace}  \def\dataAuthorsGenderUApprox{25\,M\xspace}
\def\dataAuthorsLocCctld{\num{562314}\xspace} \def\dataAuthorsLocUniv{\num{112242}\xspace} \def\dataAuthorsLoc{\num{626456}\xspace} 

\begin{document}
\title{The Impact of the COVID-19 Pandemic on Women's Contribution to Public Code}
\titlerunning{The Impact of COVID-19 on Women's Contribution to Public Code}
\date{}

\author{
  Annalí Casanueva
  \and Davide Rossi
  \and Stefano Zacchiroli
  \and Théo Zimmermann
}
\institute{
  A.~Casanueva
  \at Ifo institute, Big Data Junior Research Group, Munich. Germany.\newline
  \email{casanuevaartis@ifo.de}
  \and
  D.~Rossi
  \at University of Bologna, Bologna, Italy.\newline
  \email{daviderossi@unibo.it}
  \and
  S.~Zacchiroli
  \at LTCI, Télécom Paris, Institut Polytechnique de Paris, Palaiseau, France.\newline
  \email{stefano.zacchiroli@telecom-paris.fr}
  \and
  T.~Zimmermann
  \at LTCI, Télécom Paris, Institut Polytechnique de Paris, Palaiseau, France.\newline
  \email{theo.zimmermann@telecom-paris.fr}
}

\maketitle

\begin{abstract}

  Despite its promise of openness and inclusiveness, the development of free and open source software (FOSS) remains significantly unbalanced in terms of gender representation among contributors.
  To assist open source project maintainers and communities in addressing this imbalance, it is crucial to understand the \emph{causes} of this inequality.

  In this study, we aim to establish how the COVID-19 pandemic has influenced the ability of women to contribute to public code.
  To do so, we use the Software Heritage archive, which holds the largest dataset of commits to public code, and the difference in differences (DID) methodology from econometrics that enables the derivation of causality from historical data.

  Our findings show that the COVID-19 pandemic has disproportionately impacted women's ability to contribute to the development of public code, relatively to men.
  Further, our observations of specific contributor subgroups indicate that COVID-19 particularly affected women hobbyists, identified using contribution patterns and email address domains.

\end{abstract}

\section{Introduction}
\label{sec:intro}

Women have historically been, and remain to this day, underrepresented in free/open source software (FOSS) development~\cite{Trinkenreich2022}.
The phenomenon is not uncommon in broader contexts like STEM (Science, Technology, Engineering, and Mathematics) disciplines~\cite{hill2010whysofew, wang2017genderstem, unesco2017genderstem, reinking2018genderstem, botella2019genderstem} and computing~\cite{margolis2002womencs, frenkel1990womencs}, but it is particularly severe in FOSS, where women contributions have repeatedly been observed to be as low as 5--10\%~\cite{Trinkenreich2022}, in terms of both numbers of commits and active participants.
Trends over the past decade have been moving towards increased women participation in FOSS~\cite{ieee-sw-gender-swh}---and, more generally, in \emph{public code}, to refer to all code developed on public collaborative development platforms, no matter what the software license is---but the bottom line remains tilted towards very low women participation.

Both the negative effects of such a gender gap and needed actions to counter it have been explored in previous work; see Trinkenreich et al.~\cite{trinkenreich2022women-foss-talk} for a high-level overview of where we stand at the time of writing.
Generally speaking, actions to counter the gender gap and encourage women participation in FOSS can happen at different levels: locally within specific FOSS communities (e.g., adopting a code of conduct to prevent harassment) and more globally (e.g., programs like Outreachy or Girls Can Code as well as state-level policies).
No matter the policy level, actions need to be based on a solid understanding of the \emph{causes} of the low participation of women in FOSS.

With this study we aim to contribute to such understanding by exploring the impact of the COVID-19 pandemic---which started in early 2020 and led to lockdowns and other stringency measures around the world---on women contributions to public code.

Although major shocks like the COVID-19 pandemic are beyond the control of individual FOSS projects, they can still have a significant impact on women's participation in these projects. It is therefore essential that FOSS projects that take measures to increase the inclusion of women are aware of the impact of such external shocks, and \emph{can quantify their effect}, as part of a more encompassing monitoring of the effectiveness of their initiatives.
Furthermore, by leveraging their comprehension of external factors contributing to inequality, FOSS projects can foster a culture that acknowledges and accounts for these dynamics, e.g., when implementing meritocratic principles to determine who to promote in the community.
At the policy level, understanding the impact of the COVID-19 pandemic or similar future shocks will allow directing funds and actions towards the most effective interventions.

In fields other than computing, it has been observed~\cite{Adams-Prassl2020, Collins2021, DelBoca2020, Sevilla2020, Zamarro2021, Dang2021, Farre2020, Reichelt2021, Petts2021, Fukai2023} (see Section~\ref{sec:related-covid-gender} for a detailed breakdown of these related works) how COVID-19 correlated with and/or heightened gender disparities, due to both childcare responsibilities (that tend to fall more on mothers' shoulders) and workplace dynamics.
In computing, the impact of COVID-19 on software development has been explored~\cite{Ralph2020, Klotzman2021, Ford2021, McDermott2021, Russo2021, DaMotaSilveiraNeto2022, Pejic2023, Bao2022} (see Section~\ref{sec:related-covid-sweng} for a detailed breakdown), focusing for the most part on the effects of the pandemics on developer productivity, work patterns, and well-being.
For public code contributions, a recent study~\cite{icse-seis-2022-gender} observed a first-time decrease in women contributions in 2020 and interpreted it as possibly explained by the COVID-19 pandemic, without attempting to properly establish a causal relationship between the two.
It is hence still unknown whether the arrival of the COVID-19 pandemic \emph{directly caused} an impairment in the ability of women to contribute to public code, relatively to men.

\subsection{On the importance of establishing causality}

In this work, we aim at going beyond observing correlations.
Documenting a correlation differs from identifying a causal relationship and taking one for the other could lead to a false interpretation of the situation and ultimately to creating ineffective (if not harmful) policies.

Let us consider a simplified example that could apply to our case of study. Imagine the world is composed of two countries: country A and country B. In country A, the state has not enough budget to provide daycare facilities for children or dependent persons.
As a consequence, care responsibilities are handled by the family and, in particular, by women.
These responsibilities take time and impede women's contributions to public code. 
Country B, on the other hand, has a good state budget and is able to provide public services that take the burden of care responsibilities.
In country B, women have more time to contribute to public code.  
When the COVID-19 pandemic hits, country B is able to provide people with subsidies that allow them to stop working and make it easier to respect the social distancing measures.
On the contrary, country A has no budget to provide subsidies, people have a higher need to work and as a result are less able to avoid close interactions. 
This leads to a higher incidence of COVID-19 in country A.  

Looking at the data of these two countries, we will find a correlation between higher incidence of COVID-19 and lower contribution of women to public code.
In this case, a naive observer could think that the COVID-19 pandemic was the cause of the lower contribution of women to public code, when in fact there is a third factor (a hidden variable or confounder) that is the real cause of both the higher incidence of COVID-19 and the lower contribution of women to public code: state budget.

Now consider a private foundation dedicated to supporting open source, and which has as one of its missions to foster gender diversity in open source projects. When the next pandemic hits, this foundation, basing its decisions on the naive conclusion from observed correlations, wants to avoid a negative impact on gender diversity in public code, and they may be inclined to redirect some of their funding to distributing masks to the public.
In this fictional example, this strategy will turn out to be completely ineffective for its objective, since the real cause of the lower contribution of women was state budget, and not the pandemic.
Had they known the real cause of the lower contribution of women, they could have used their resources more effectively, e.g., by supporting measures that relieve women from care responsibilities.

\smallskip

Beyond identifying actual \emph{causes} affecting women's ability to contribute to public code, \emph{measuring  the impact} of each cause is also key.  When resources for an intervention are limited, knowing which factors have a greater impact, allows to allocate resources more effectively. Alternatively, when several factors play a role simultaneously, knowing the impact of each one allows to determine if a specific intervention is effective or not and permits to truly address the problem at hand.

Imagine for example that a well-intentioned group of software developers implemented a code of conduct in the community with the goal that women feel more welcome and are able to contribute more. They implemented this code of conduct during the pandemic. To evaluate if their intervention had an effect, they measured the relative level of activity of women before and some weeks after the intervention. They found out, to their deception, that after the implementation of their new code of conduct, women in fact contributed relatively \emph{less} that before. Being able to factor out the direct effect of COVID-19 would have made them realize that in fact their policy was \emph{effective} as the decrease in the relative level of contribution was in fact lower than what would have happened without the intervention. The culprit to the observed decrease in relative contribution is COVID-19 and the code of conduct was, in fact, able to counter-balance its effect a little.
\smallskip

Building a body of empirical literature that is able to \emph{identify the causes} that prevent women from contributing to public code in larger numbers and \emph{quantitatively measure} them is key, particularly in a field that aims to identify interventions that could help to effectively increase diversity. 

\subsection{Contributions}

\paragraph{Research questions.}

This work explores the general theme of: what were the effects of the COVID-19 pandemic on gender diversity in contributions to public code?
More precisely, we will answer the following research questions:
\begin{itemize}

\item \textbf{\RQcause} \emph{How did the COVID-19 pandemic impact the ability of women (relatively to men) to contribute to public code?}

\item \textbf{\RQwho} \emph{Which groups of women contributing to public code were impacted the most (relatively to men in the same groups), in their ability to contribute, by the COVID-19 pandemic?}

\end{itemize}

\paragraph{Methodology.}

In order to answer these research questions, we start from a large-scale dataset obtained from the Software Heritage archive~\cite{swh-msr2019-dataset} and comprised of \dataCommitsInitApprox (billion) commits contributed by \dataAuthorsInitApprox (million) authors via major software development forges (e.g., GitHub, GitLab.com) and source code distribution platforms (e.g., package manager repositories), over a period of several decades.
We then, on the one hand, detect the gender of contributors using name-based heuristics and, on the other hand, geolocate the same population to individual countries around the world by applying heuristics based on the domain of contributor emails as recorded in commit data.
(See Section~\ref{sec:dataset} for details.)

Data obtained this way allows observing commits to public code before and during the pandemic and to aggregate them by country, period, and contributor groups.
It also allows studying the track record of contributors via features like: when they started contributing, in which years they were active, and other time-based work patterns.

We complement large-scale public code data with world population and COVID-19 data, and most notably: worldwide daily COVID-19 deaths reported to WHO (World Health Organization), global population data from the UN (United Nations), and the Oxford COVID-19 dataset about the level of stringency of governments responses to COVID-19.

Then, we answer the stated research questions applying \emph{difference in differences} (DID)~\cite{angrist2009mostly}, a technique from econometrics that allows us to unveil causality, rather than mere correlation, between observed changes in selected variables, such as being a woman or having contributed to public code in a given period and living in a country where people were more exposed to COVID-19 restrictions.
(See Section~\ref{sec:methodology} for details.)

\paragraph{Key findings.}

We answer \RQcause by showing that higher exposure to COVID-19 caused a statistically significant reduction in contributing activity (as measured in number of commits per week) by women, relatively to men.
The result is robust and holds for two different measures of exposure to COVID-19: COVID-19 deaths \emph{per capita} in a given country and strictness of anti-COVID-19 measures.
For the latter COVID-19 measure, our result also holds when changing the measure of contributing activity from number of commits to number of active days (with at least one commit) per week.
In particular, one additional COVID-19 related death per \num{100000} inhabitants widens the gap of the number of contributions between men and women by 2.1\%. Alternatively, passing from a situation with no restrictions to a situation with social distance measures of 75\% of the maximum of our measure, increases the gap by around 28\%.

Regarding \RQwho, we explore four different ways of categorizing contributors. Two of these categorizations aim to distinguish between hobbyists and paid contributors. Both of them lead us to the same conclusion: the relative impact of COVID-19 on women is stronger for hobbyists, as identified by contributing mostly outside of working hours and by using non-professional email addresses.

\subsection{Paper structure}

Section~\ref{sec:related} reviews related work on COVID-19, gender issues, and software development.
Section~\ref{sec:dataset} presents the datasets we used and how we treated and augmented them, including gender detection and geolocation.
The use of difference in differences is discussed in Section~\ref{sec:methodology}.
Section~\ref{sec:results} presents obtained results, answering the stated research questions.
The interpretation of our findings and threats to their validity are discussed in Section~\ref{sec:discussion}.
Section~\ref{sec:conclusion} concludes the paper and suggests leads for future work.

\subsection{Data availability}
A replication package for the work presented in this paper is available~\cite{this-replication-package}.
See Section~\ref{sec:threats-reliability} for more information about what it includes.

\section{Related work}
\label{sec:related}

The effects of the COVID-19 pandemic on various organizations have been extensively investigated and analyzed, delving into the multifaceted ways in which these events have influenced the dynamics of involved entities~\cite{Donthu2020, Kaushik2020}.
The present work sits at the crossroad between: (1) gender issues, (2) the COVID-19 pandemic, and (3) software engineering in society. 
While these three topics \emph{together} have only been studied tangentially (see~\cite{Machado2021} for an exception), various intersections among them have been analyzed.
In this section, we review numerous related works in this broad space, organizing them based on the possible overlaps of the three topics.

\subsection{COVID-19 and gender issues}
\label{sec:related-covid-gender}

A first body of related work looked at the potential of the COVID-19 crisis
to amplify gender disparities in both workplace dynamics and household management tasks (houseworks, children and elderly care).
Most of these studies operate on data elicited via online surveys and adopt regression analysis to assess the correlation between the pandemic and gender disparities.
Apart from some minor methodological differences, what differentiates these studies is the geographical origin of the respondents. 
Most studies focus on a specific nation/geographic area, with most findings suggesting that emerging phenomena are common across America, Europe, and Japan.

\paragraph{Takeaways.}
Existing studies on the impact of COVID-19 on the gender gap, reviewed in details below, establish correlation rather than causation (with a few exceptions, like~\cite{Fukai2023}) between the two.
This marks a major difference with respect to this paper, where we employ difference in differences (DID) to establish that changes in public code contributions by women are indeed \emph{due to} COVID-19.

\paragraph{Study details.}

Adams-Prassal et al.~\cite{Adams-Prassl2020} use real-time survey data, gathered in the US, UK, and Germany in April 2020, to examine changes in income, employment, mental health, and care work among different groups of people during the pandemic.
The paper highlights that while everyone has been affected by COVID-19, there are significant differences across socioeconomic groups.
Authors also find that women have been disproportionately affected by the pandemic due to their higher care burden, both before and during the crisis.
For example, they found that in the US and the UK, women were 6.5\% and 4.8\% more likely to lose their jobs.

Collins et al.~\cite{Collins2021} use panel data from the US Current Population Survey
to examine changes in mothers’ and fathers’ work hours from February through April 2020. 
Using fixed effects models they found that mothers with young children have reduced their work hours 4--5 times more than fathers.

Del Boca et al.~\cite{DelBoca2020} employ survey data gathered in April 2020 from a representative sample of Italian women. It seeks to examine the effects of altered working arrangements due to COVID-19 on housework, childcare, and homeschooling within couples where both partners work. 
Authors discovered that, except for women working at their usual workplace, all surveyed women spent more time on housework. 
Men's housework time depends on their partners' working arrangements, increasing when partners work away. 
Childcare time is symmetrically affected, with both women and men spending less time if working away. 
Working women with young children find work-life balance challenging during COVID-19, especially if their partners continue to work outside the home.

Sevilla et al.~\cite{Sevilla2020} present a regression analysis of 2782 responses to the Ipsos MORI omnibus survey of May 2020, provided by respondents living in the UK, aged 18--60, part of families with children under the age of 12.
The authors found that the increase in childcare hours for women is less influenced by their employment status compared to men. 
This often results in many women managing both work and a significantly increased childcare burden, potentially leading to adverse effects on their mental health and future career prospects. 
Notably, households where men were not employed show greater strides toward achieving a more equal distribution of childcare responsibilities.

Zamarro et al.~\cite{Zamarro2021} analyzed approximately 7000 respondents to the Understanding Coronavirus in America Tracking Survey in March 2020 to understand how fathers and mothers were coping with this crisis in terms of childcare provision, employment, working arrangements, and psychological distress levels.
They found that women shouldered a greater childcare burden than men during the COVID-19 crisis, even while working.
The current working situations of mothers seem to have minimal impact on their childcare responsibilities.
However, this division of childcare is linked to reduced working hours and an increased likelihood of working mothers transitioning out of employment.

Dang et al.~\cite{Dang2021} analyzed 6089 responses to surveys collected across 6 countries including China, South Korea, Japan, Italy, the United Kingdom and the four largest states in the United States, allowing an analysis that is not specific to a single country's dynamics.
The survey contained questions on fundamental demographic factors of participants, their employment and living arrangements, health status and diseases, self-reported economic and non-economic consequences of the pandemic, behaviors, beliefs about the pandemic, and evaluations of government responses (parts of a dataset described in~\cite{Belot2020}).
The regression analysis conducted by the authors indicate that women face a 24 percent higher likelihood of experiencing permanent job loss due to the outbreak compared to men.
Additionally, women anticipate a 50 percent greater reduction in their labor income compared to men.

Farre et al.~\cite{Farre2020} analyzed 5001 responses to a survey distributed in Spain to a population aged 24-50 in May 2000.
The questions focused on changes in employment during the lockdown and changes in the distribution of childcare and housework. 
The authors' analysis reveals that nearly 20\% of workers, particularly those with lower education levels, experienced furloughs. 
Temporary job losses slightly affected women more than men, and women were more likely to work from home during confinement. 
The study also found a significant increase in childcare and housework responsibilities for parents, with women continuing to bear the majority of the burden. 
On average, there is a gender gap of about 17\% in parents' shares of childcare and housework during the lockdown.

Reichelt et al.~\cite{Reichelt2021} collected 5008 responses from a survey conducted in May--June 2020 in the USA, Germany and Singapore. 
The panel included individuals working either full- or part-time in January 2020, providing retrospective responses about their employment status and couples where both the respondent and their cohabiting partner worked full- or part-time in January, offering retrospective information about themselves and their partners. 
Their regression analyses indicate that women experienced more frequent transitions to unemployment, reductions in working hours, and shifts to working from home compared to men, although the extent varied across the three countries studied. 
Additionally, among couples initially employed at the start of the pandemic, men tended to express more egalitarian gender-role attitudes when they became unemployed while their partners remained employed. 
In contrast, women expressed more traditional attitudes when they became unemployed and their partners remained employed.

Petts et al.~\cite{Petts2021} used data from U.S. parents (623 mothers and 891 fathers) to empirically investigate whether the challenges posed by the loss of childcare and new homeschooling demands are linked to employment outcomes early in the pandemic.
They also explore whether the pre-pandemic division of child care is associated with parents' employment.
The authors find, by the means of regression studies, that for parents with young children, the loss of full-time childcare increased the risk of unemployment for mothers but not for fathers. 
However, greater father involvement in childcare helped mitigate negative employment outcomes for mothers of young children.
In the case of parents with school-age children, participation in homeschooling was associated with adverse employment outcomes for mothers but not for fathers.

Fukai et al.~\cite{Fukai2023} used data from the Labor Force Survey, published by Japan’s Ministry of Internal Affairs and Communications, encompassing about \num{40 000} monthly responses dating back to well before the onset of the pandemic. 
This allowed the study to adopt a regression discontinuity design (RDD) approach and estimate that the pandemic decreased (causation) the mothers' employment rate by about 4\%, mostly because of increased childcare responsibilities.

Other relevant works~\cite{Power2020, Blundell2020} do not offer analyses of new data; instead, they discuss the results presented in previous works, aiming to uncover underlying societal issues and propose policies addressing gender inequality.

\subsection{COVID-19 and software development}
\label{sec:related-covid-sweng}

Professional software development teams, as well as online open source communities, also attracted the attention of researchers.
A second body of relevant literature follows the tradition of empirical software engineering, utilizing both elicited and mined data, offering analyses that encompass both quantitative and qualitative aspects.

\paragraph{Takeaways.}
Overall, empirical studies on COVID-19 on software development, open source or otherwise, have found various sorts of correlations between the two, ranging from productivity changes to work patterns (work hours and weekdays) and from what activities are performed (e.g., development vs documentation writing) to developers' well-being.
We complement this body of work by showing that COVID-19 has caused, rather than merely being associated with, a decrease in the ability of women to contribute to public code, providing a detailed analysis of which groups of women from the pre-pandemic population were affected.

\paragraph{Study details.}

Ralph et al.~\cite{Ralph2020} examined 2225 questionnaires to assess the effects of the pandemic on software developers.
The study adopts non-parametric hypothesis tests and structural equation modeling to establish a negative correlation between developers working from home because of social restrictions and their well-being and productivity.

Klotzman et al.~\cite{Klotzman2021} used data mined from GitHub and Stack Overflow to determine if COVID-19 impacted open source development.
The authors could not find conclusive indications about GitHub, but they found out that there has been a significant increase in the number of new users and questions posted on Stack Overflow during the first months of the pandemic (March/April 2020).

Ford et al.~\cite{Ford2021} analyzed a couple of online surveys, one with 1369 responses and another with 2265 responses, distributed to two sets of workers (the first in King County, WA, USA; the second in all USA states) among software engineers at Microsoft. 
The surveys were conducted from March to May 2020, when the company had support in place for its developers and remote work.
The authors discuss aspects such as productivity, benefits, challenges, and recommendations that emerged from the surveys.
They found a dichotomy in developer experiences influenced by various factors: what is perceived as a benefit for some is considered a challenge by others.
For instance, the necessity of staying at home is seen as a benefit for being close to family members but as a challenge for those struggling with working space and dealing with interruptions.
The analysis of the responses not only addresses the different perceptions of the implications of the pandemic, but also provides deeper insights into the narratives stemming from these diverse experiences.

McDermott et al.~\cite{McDermott2021} extracted data from the GitHub timeline to assess changes in work patterns potentially associated with the advent of the pandemic. 
Their work shows that developers were more active on weekends and outside of regular work hours than they were in previous years. 
It also detects that COVID-19 was associated with a jump in overall activity: during the early lockdown period, GitHub users were not merely reallocating a fixed budget of work hours; instead, they were actively increasing their overall work hours.
Self-reported men and women users exhibited comparable patterns of increased working hours shortly after March 2020. 
However, men responded more promptly and consistently to the challenges posed by the pandemic. 
In contrast, women were slower in reallocating work, and their response was more inconsistent.

Russo et al.~\cite{Russo2021} describe a two-wave longitudinal study involving almost 200 selected software professionals on a global scale.
Aim of the study was to infer pandemic-induced changes in everyday tasks correlated with perceived well-being, productivity, and other pertinent psychological and social factors.
The findings presented by the authors indicate that the time allocated to specific activities from home was comparable to when working in the office.
However, the duration developers devoted to each activity showed no correlation with their well-being, perceived productivity, and other variables.

Da Mota Silveira Neto et al.~\cite{DaMotaSilveiraNeto2022} combined a mining approach (conducted on 100 GitHub Java projects) with the analysis of 279 survey responses. 
Various code and project metrics were used to assess productivity, while the surveys aimed to evaluate projects and developers' well-being. 
In line with other previously discussed studies, the authors found that the impact of COVID-19 is not a strict dichotomy (reducing vs.~increasing productivity); rather there exists a spectrum, where significant proportions of respondents hold differing opinions on the matter.

Pejić et al.~\cite{Pejic2023} analyzed the GitHub timeline, comparing the event patterns in 2017-2019 and comparing them with the period 2020-2022.
They observed that, overall, events associated with individual development have either maintained or experienced an increase in their trends.
In contrast, events linked to community activity (such as forking) or contributions to documentation have exhibited a slight decrease.

Bao et al.~\cite{Bao2022} analyzed developers’ daily activities (commits, code reviews, build, etc.) from Baidu Inc., one of China's largest IT companies.
The authors aimed to evaluate the impact of the pandemic-mandated work-from-home arrangement on productivity. Similar to other studies, they identify both positive and negative impacts

\subsection{Software development and gender issues}
\label{sec:related-sweng-gender}

Leaving COVID-19 aside, a third and final body of literature have analyzed in recent years, gender-related issues within software development teams.
A comprehensive reference that analyzes the most relevant works on this topic is Trinkenreich et al.~\cite{Trinkenreich2022}, which provides a framework to position the various contributions and to which we refer the interested reader.

Usually studies operate on elicited data (mostly online surveys), mined data, or both.
Those operating on mined data extract various types of items, using them as traces of productivity and/or collaboration: commits, issues, pull requests, code reviews, but also social software-like elements, such as GitHub ``stars''.

While most of the presented works mainly aim at developers operating on FOSS/public code projects (all, among those adopting software mining techniques), there are notable exceptions.
For example, \cite{Trinkenreich2022a} is a recent study based on the analysis of a survey (94 respondents) conducted among Ericsson’s female workforce, allowing the authors to identify and assess challenges (maternal wall, glass ceiling, impostor syndrome, etc.) and propose associated mitigations (support work-life balance, empower women, support women’s career growth, etc.).

\paragraph{Takeaways.}
The overall picture painted by these works is disheartening: under-representations and biases against women feed each other in an endless loop. 
The level of awareness, however, is rising; a few strategies to ease (and retain) women’s participation are becoming more common (like the very elementary adoption of explicit codes of conduct), and recent positive trends have been observed~\cite{Prana2022, icse-seis-2022-gender}, until the arrival of the pandemics.

The present work aims at completing the picture of the current state of gender issues in software development, by looking at the intersection of public code, COVID-19, and the causal relation among the two.

\section{Dataset}
\label{sec:dataset}

To answer the stated research questions, we need data about public code production over time, as well as geolocated data about the COVID-19 pandemic.
In this section, we describe the datasets we obtained to address these needs and how we preprocessed them before further analysis.
The rest of our methodology pipeline, from the dataset on, is detailed in Section~\ref{sec:methodology}.

\subsection{Public code data}
\label{sec:dataset-code}

\paragraph{Initial dataset.}
As our starting point for public code data we obtained from the Software Heritage initiative~\cite{swh-msr2019-dataset} a dump of all the commits archived by the project up to March 2024.
Software Heritage~\cite{cacm-2018-software-heritage} is the largest archive of source code and its development history, as captured by modern version control systems (VCS), like Git.
It archives both code and commit data from hundreds of million projects developed on, or distributed from, major development forges (GitHub, GitLab, Bitbucket, etc.), free/open source software (FOSS) distributions (Debian, GNU, Nix, etc.), and package manager repositories (PyPI, NPM, Maven, etc.).
The specific data dump version we obtained contains \dataCommitsInit commits (unique by commit SHA1, computed in the same way Git does), authored by \dataAuthorsInit authors (unique by full name strings, as recorded in VCSs).
The dataset does not contain the actual files (source code, data, etc.) associated to each commit, but rather all \emph{metadata}, such as authors, timestamps, commit messages, etc.

Obtained commits came as two relational tables, one for commits and one for authors.
Each row in the commit table contains the following columns: commit SHA1 identifier, author and committer timestamps, author and committer identifiers. Each of the last two columns references the author table via a foreign key.
The distinction between authors and committers comes from Git, which allows a code integrator to commit a change authored by someone else.
For this study, we used authors and ignored committers, because the difference between the two is immaterial for our research questions and negligible in amount.
Each row in the author table contains two fields: full name and email, parsed (by Software Heritage) from the raw author identification strings stored natively by VCSs in the form \verb+"Author Name <foo@example.com>"+.

\begin{figure}
    \centering
    \includegraphics[width=\textwidth,trim=15mm 15mm 15mm 15mm]{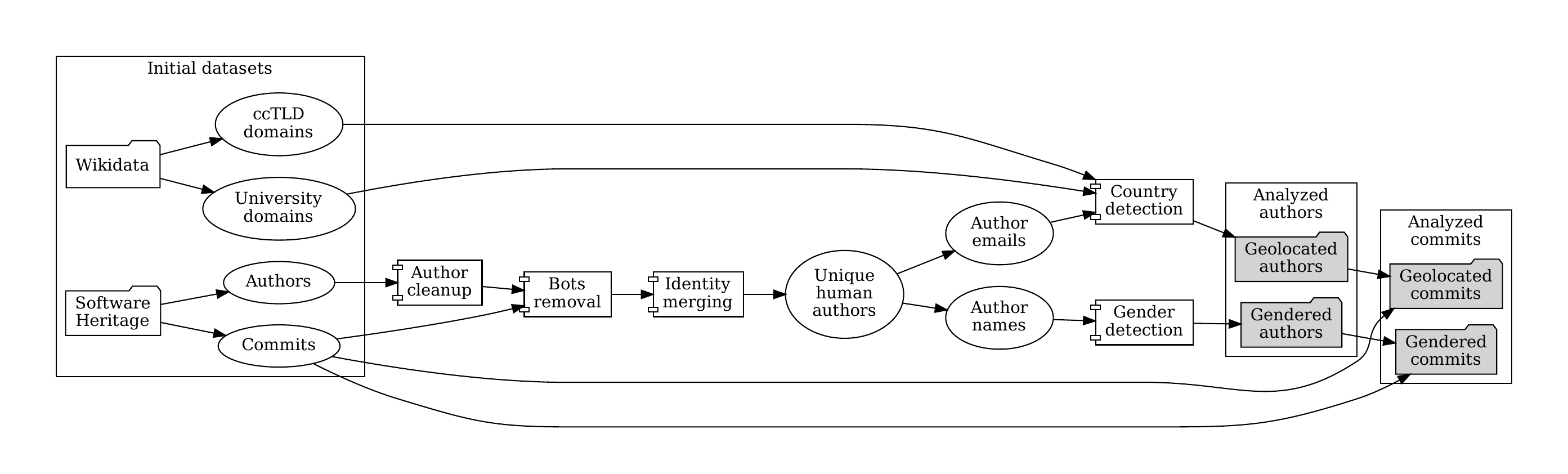}
    \caption{Public code data used in this study. Starting from commits archived by Software Heritage, we split them into author and commit data, classifying each class along two axes: gender and country of origin.}
    \label{fig:dataset-swh}
\end{figure}

\smallskip

From this point on, the preprocessing steps applied to public code data are those depicted in Figure~\ref{fig:dataset-swh} and described in the following.
Some of them---Author cleanup and Gender detection---have been implemented, with specific deviations described below, by replaying the replication package of Rossi and Zacchiroli~\cite{icse-seis-2022-gender} on the obtained version of Software Heritage data.
Crucially, this version spans 2.5 additional years of public code data (with respect to~\cite{icse-seis-2022-gender}), encompassing the arrival of the COVID-19 pandemic and accompanying stringency measures.
However, we also learned that there are significant delays in the Software Heritage archiving of some public repositories, which could bias the resulting data analysis,\footnote{Active repositories are archived every two weeks on average. However, new repositories are archived in a different pipeline, which prioritizes recently updated repositories, leading to an archiving delay that can grow up to two years for new, but inactive, repositories.} therefore we only consider commits up to May 2022 in our analysis, for which we have exhaustivity guarantees despite the archiving delay.
Other parts of the data processing pipeline were implemented from scratch, as detailed below.

\paragraph{Data cleanup.}
The \emph{author cleanup} step removes from the initial dataset authors whose names are either unusable for practical reasons, or implausible as names of real people.
Specifically, this step drops: names that cannot be decoded as UTF-8 strings, email addresses used instead of (or in addition to) names, names consisting of only blank characters, names containing more than 10\% non-letter Unicode characters, and names longer than 100 characters.
About one quarter of the initial author strings were removed by this filtering step, with \dataAuthorsPostCleanup authors remaining for further analysis.

As it is customary for commit analysis practices, we further removed bots from the dataset (\emph{bots removal} in Figure~\ref{fig:dataset-swh}) and merged together authors appearing with different identities (\emph{identity merging}), in this order.
For bots removal, we used the approach and dataset of Dey et al.~\cite{dey2020botsremoval}, removing from the dataset all authors whose emails match a bot in their dataset.
After this step, \dataAuthorsPostBots authors remained.

For identity merging, we implemented the straightforward approach of merging together all authors sharing the same email address.
Also, when the same email address is used by ten or more authors we dropped them all.
This was usually due to the adoption of generic, non realistic, email addresses like \verb+you@example.com+ (38K occurrences), \verb+devnull@localhost+ (12K occurrences), and so forth.
After this step, \dataAuthorsPostMerge authors remained, referred to as \emph{Unique human authors} in Figure~\ref{fig:dataset-swh}.

\paragraph{Gender detection.}

To answer the stated research questions, we need to associate a gender and a country to the authors of commits in our dataset.
Some preliminary considerations are in order before attacking the first task, which we refer to as ``gender detection'' in the following.
At this scale (tens of million authors), manual approaches based on interviews, where individuals state what gender they identify with, are not feasible.
Hence, following in the steps of previous work at similar scales~\cite{ieee-sw-gender-swh, icse-seis-2022-gender}, we rely instead on automated mechanisms (described later) that detect the gender of authors based on their public names, as recorded in code commits.
Those tools are generally restricted to binary gender assignment.
With that, we do not intend to arbitrarily define people within a binary gender confinement regardless of their preferences and sensitivity.
In fact, none of the gender-related decisions made by the automated techniques used in this paper make sense when applied to \emph{individuals} present in the analyzed corpus.
The meaning of the exercise is statistical in nature and aims only to address the stated research questions.
The used approach makes sense only in aggregate form and carries with it the unavoidable limitations that name-based gender detection entail.
We elaborate further on those limitations in Section~\ref{sec:threats}.

To detect the gender of an author from our public code dataset, we apply \PKGGG~\cite{genderguesser} to its full name, as recorded in a commit.
\PKGGG is an open source Python library for gender detection, based on first name frequencies around the world, which is often used in related work.
Our choice of \PKGGG over alternatives is based on the fact that it is open source, which helps with study repeatability, and an actual self-hostable tool, rather than a remote service API which would be unusable (due to costs or execution time) at this scale.
Furthermore, Santamaria et al.~\cite{santamaria2018genderapi} conducted a comparative benchmark of \PKGGG and its main competitors, including payware commercial API services, showing that even if \PKGGG is not the best tool in absolute terms, it works comparatively well with geographically diverse datasets, as ours is by construction (it contains commits coming from all over the world).

As in previous works based on similar datasets and \PKGGG~\cite{ieee-sw-gender-swh, icse-seis-2022-gender}, we address the issue that \PKGGG works primarily on first (given) names by applying a \emph{majority criterion} for the final gender determination.
We first tokenize full name strings by splitting at each blank, hyphen, or case change (to also address CamelCase notation, which is used by several authors in the dataset), and then use \PKGGG to determine the gender of each token. 
Among the tokens that were associated with a gender, if and only if a \emph{strict majority} of name tokens for a given author full name is detected as belonging to one gender, we associate the majority gender to the author; otherwise their gender will remain unknown.

Differently from Rossi and Zacchiroli~\cite{icse-seis-2022-gender}, due to identity merging we reach this data processing step with multiple author names, each composed by multiple tokens, associated to each unique author.
To associate a gender to each \emph{unique} author, we first applied the majority criterion token by token to each identity, and then applied it again on the genders obtained for each identity to determine the final gender of each unique author.

\smallskip
In the end, we identified a total of \dataAuthorsGenderM men contributors and \dataAuthorsGenderW women contributors, counted as unique authors after identity merging.
A gender could not be identified automatically for the remaining unique authors in the dataset (about \dataAuthorsGenderUApprox of them), which were excluded from further analysis.

\paragraph{Geo-localization.}
In order to quantitatively compare the effects of COVID-19 on contribution to public code around the world, we need to associate commit authors in the dataset to a geographic position, or ``geolocate'' them.
More specifically, due to the granularity of COVID-19 data, we aim at geolocating commit authors to specific world \emph{countries}.

As for gender detection, the scale of our data rules out manual approaches like interviewing contributors to ask where they are from---we need an automated country detection approach.
We based ours on the email domain, which is recorded as part of commit metadata in our dataset.

We look up the email domain of each author in two datasets: country code top-level domain (ccTLD) and university websites, both retrieved from Wikipedia/Wikidata.
The ccTLD dataset\footnote{A human-readable version of the ccTLD dataset can be found at \url{https://en.wikipedia.org/wiki/Country_code_top-level_domain}, accessed November 2023. Note that we decided to discard a few ccTLDs that we considered problematic because they are widely used outside the country that they represent: .ai, .cc, .co, .io, .me, .sh, .tv, .vg, .yt (used in so-called ``domain hacks''). We also excluded .gp because it is associated with several ISO3 country codes.} associates top-level Internet domains (e.g., \texttt{.fr}, \texttt{.it}, \texttt{.mx}) to specific countries (e.g., France, Italy, and Mexico, respectively).
The university website dataset associates full website URLs (such as \texttt{https://www.telecom-paris.fr/}, \texttt{https://www.unibo.it}, or \linebreak \texttt{https://www.unam.mx}) to universities around the world, and the country where the university is incorporated.
To obtain this dataset, we queried the Wikidata Query Service with a SPARQL query\footnote{The query can be found in the replication package of this paper~\cite{this-replication-package}.} returning all known universities on Wikipedia and, for each of them, its website and country of incorporation (both are available as Wikidata properties on university entities).

To associate authors to countries, we extract their location from their ccTLD and also match their email domains to university websites (stripping common and non-meaningful for us prefixes like URI schemes and \texttt{www.}).
There are very few cases where a university domain is associated with a country that is different from the one associated with the ccTLD; we discard the authors in those cases.
In case of failure of both lookups, we leave the authors non-geolocated and discard them from further analysis.

Because we were interested in matching authors to companies to answer \RQwho, we also used Wikidata to retrieve a list of companies, their websites, and their countries of incorporation. We do not use this dataset to geo-localize authors, because a consistency check between the declared location of the company and the time zone of the majority of commits by the author (following the same method as described below) showed that this geo-localization method was less reliable than the ccTLD-based and university-based methods.
However, we used this dataset to filter out authors whose email domain was associated with a multinational corporation, because we could not reliably associate them to a single country, or authors for which the ccTLD and reported company location were inconsistent.

\medskip

Finally, for the present study, we were only interested in authors who created at least one commit in 2019, the year before the pandemic (because we wish to measure the impact of COVID-19 on preexisting contributors). Excluding authors without any contribution in 2019, and authors which we could not geo-localize, we obtain \textbf{a final set of \dataAuthorsLoc unique human contributors to public code, for which both gender and geographic location at country granularity were identified}, out of which \dataAuthorsLocUniv were located via their university and \dataAuthorsLocCctld via their ccTLD.

\medskip

We validated the accuracy of our geolocation mechanism, by relying on the latest available dataset from the (now-defunct) GHTorrent project~\cite{GHTorrent} as ground truth.
The \texttt{User} table in GHtorrent includes both an \texttt{email} and a \texttt{country\_code} field; the latter being an ISO 3166-1 alpha-2 country code.

The country code is obtained by GHTorrent applying Named Entity Recognition techniques to the \texttt{location} attribute self-declared by GitHub users on their profiles.
Although the possibility exists that users provide fictitious or inaccurate locations, we contend that this is not the case for the vast majority.
To support this assertion, we examined the commit timestamps of all users with a non-empty \texttt{country\_code}, verifying whether their time zone offset is compatible with the associated country's time zone offset at commit time.
Our findings show that more than 90\% of the commits are compatible with the declared country, assuming a reasonable amount of anomalies (e.g., developers being temporarily on a trip in a different country).
This figure gives us confidence that the vast majority of self-declared locations are correct, and hence that it is reasonable to rely on GHTorrent location as ground truth for validation.
(Note that GHTorrent is only useful for us as validation, rather than as a source for geolocating authors, because it is both limited to GitHub and largely outdated, missing a large proportion of our authors.)

Subsequently, we applied our geo-localization mechanism to the set of developers that have country information in the GHTorrent dataset, comparing our results with the \texttt{country\_code} there.
Out of the initial \num{1 127 255} entries, the classification successfully identified a country for \num{108 318} (9.6\%) authors.
Among the localized entries, the accuracy was 0.88, with a weighted mean precision of 0.91 and a weighted mean F1 score of 0.88.
In the top 29 countries by support (number of entries in the validation set), accounting for the 90th percentile, only four entries had F1 score $<0.85$.

\paragraph{Number of commits per week.}

For each gendered and geolocated unique human author, we count the number of commits authored each week (from Monday to Sunday), distinguishing in particular commits during working hours (defined as 7.00\,AM--6.59\,PM Monday to Friday) and outside working hours.
We include in our dataset all commits made between Monday, May 28th, 2018 and Sunday, May 29th, 2022, to include about four full years of contribution activity data (209 complete weeks), including more than a full year of data before the onset of COVID-19, and without using any data for which the archiving by Software Heritage was not guaranteed to be complete.
Note that in the dataset descriptive statistics (Table~\ref{descriptives}), the minimum number of commits in 2019 is 0 despite only considering authors that contributed at least once in 2019, because the aggregation at week level may lead to commits performed on the last few days of a year to be counted in the following year.

\paragraph{Number of active days per week.}

In addition to the number of commits, we also count the number of active days per week, defined as the number of days in the week in which the author authored at least one commit. This measure is used as an alternative specification of the dependent variable when answering \RQcause.

\subsection{COVID-19 and demographic data}
\label{sec:dataset-covid}

\label{sec:covid-data}

\paragraph{COVID-19 deaths.}
To know about weekly COVID-19 attributed deaths, we used the dataset ``Daily cases and deaths by date reported to WHO'' from the World Health Organization (WHO) Coronavirus (COVID-19) Dashboard,\footnote{\url{https://covid19.who.int/data}, accessed November 2023} which aggregates COVID-19 data by country/territory and varies per day. We aggregate the data per week. 

\paragraph{Lockdown stringency.} We use data from the Oxford COVID-19 Government Response Tracker~\cite{hale2020oxford} to measure the restrictiveness of the government's pandemic policy.
Notably, we use an original stringency index which captures the strictness of lockdown-style policies.
This index reports values between 0 (no measure implemented) and 100 (highest stringency of social distance measures) and varies across countries and day.
We aggregate the data per week and re-scaled it, making it vary between 0 and 1.

\paragraph{Population.} As general population data,
we relied on the ``Demographic Indicators'' dataset,\footnote{\url{https://population.un.org/wpp/Download/Standard/MostUsed/} accessed November 2023} elaborated by the UN Department of Economic and Social Affairs Population Division.
This data includes the total population of each country as of January 1st of every year.
We divided the number of deaths per week of each year by the population on the 1st of January of the same year. 

\bigskip

Table~\ref{descriptives} shows the number of observations and basic descriptive statistics for the variables that were used in the empirical analysis.

\begin{table}
  \caption{\textbf{Descriptive statistics of main variables}
  }
 
 \begin{center}
 \label{descriptives}
 \begin{tabular}{@{}lrrrrr@{}}
 \hline
 & & & \textbf{Std.} & & \\
 \textbf{Variable} & \textbf{Observations} & \textbf{Mean} & \textbf{dev.} & \textbf{Min} & \textbf{Max} \\ \hline
 Commits per week (all) & \num{130670400}  &  0.592 &  10.53 & 0 & \num{36554} \\
 Commits per week (women) & \num{17051160} & 0.381 &   4.162 & 0 & \num{4 634} \\
 Commits per week (men) & \num{113619240} & 0.624 & 11.18 &   0  & \num{36554} \\ \hline
 Active days per week (all) & \num{130670400} & 0.150 & 0.663 & 0 & 7 \\
 Active days per week (women) & \num{17051160} & 0.106 & 0.546 & 0 & 7 \\
 Active days per week (men) & \num{113619240} & 0.156 & 0.678 & 0 & 7 \\ \hline
 COVID-19 deaths per week per & \num{78402240} &  1.702 & 2.215 & -0.489 & 103.9 \\
 \num{100000} inhabitants (2020+) & & & & & \\
 Social distancing index (2020+) &  \num{77864036} &  0.505 &  0.217 &  0 & 1 \\
 \hline
 \end{tabular}
 
 \end{center}
 Note: This table shows the basic descriptive statistics of the main variables used in the empirical analysis. The weekly number of COVID-19 related deaths per \num{100000} inhabitants includes negative values because some countries (particularly at the beginning of the pandemic when the methodology to count COVID-19 related deaths was not yet stabilized) corrected the number of reported deaths from one week to the other, leading to some weeks with reported negative deaths to compensate for over reporting in the past.  This, however, represents a small amount of country-week observations. 
 
 \end{table}

\section{Methodology}
\label{sec:methodology}

To assess the causal effect of an event, we would ideally need to compare individuals in a world where the event took place, with the same individuals at the same time in an alternative world where the event did not happen.
This is, of course, impossible.
To approach this ideal scenario, scientists instead often set up to compare two randomly selected groups that only differ to the extent that one group is exposed to the event (the treated group) and the other is not (the control group).
If the assignment to a group is random and the groups are large enough then, on average, the characteristics of the two groups are equal and the differences between them can be considered to occur only because of the exposure to the event of interest.
However, in many settings, researchers are interested in studying the causal effect of an event, the exposure to which cannot be randomized across groups.
This is the case of the COVID-19 pandemic.
In these cases, several econometric techniques can be used to determine causal effects, depending on the context and the data structure.
See the introduction of the recent paper by Graf-Vlachy and Wagner\cite{graf-vlachy_cleaning_2024} for a summary of the four main econometric techniques that can be used to determine causal effects from observational data (and for more examples of why causal inference is important in software engineering research).

In this paper, we use a variation of a difference in differences (DID) strategy~\cite{angrist2009mostly}, one of the four main techniques mentioned above, which has already found applications in software engineering research~\cite{maldeniya_herding_2020,fang_this_2022,russo_latona_shock_2024},
to estimate the causal impact of COVID-19 on the contribution of women to public code.
This strategy takes advantage of the rich panel dimension of the data, i.e., being able to follow every author throughout their full contribution history.

The basic idea of DID is to compare a treated group and a control group that don't necessarily share the same characteristics.
It is enough to assume that, in the absence of treatment, the trends for the outcome variable for the two groups would be parallel.
In this setting, one group is exposed to the event and the other is not, but the two groups are not constructed randomly and can hence have different characteristics. 
Since we can observe the difference in the outcome variable between the two groups \emph{before} the event of interest, we can subtract the difference in the outcome variable between the two groups (before the event) from the difference between the two groups after the event and obtain an estimation of how the treated group \emph{would have looked like} in the absence of the event (assuming parallel trends).
Any deviation from this estimation is considered to be the effect of the exposure to the event. 

In our case, though, we do not have a true control group that we can consider never exposed to the event, because at some point or another, the entire world has been exposed to the COVID-19 pandemic.
Therefore, we exploit the fact that different countries were exposed to COVID-19 at different points in time and with different intensities.
In the case where there are several groups and several points in time, this method is called \emph{two-way fixed effects (TWFE)}~\cite{angrist2009mostly}.

Finally, another difference in our analysis with respect to a classical DID strategy is that we are interested in the differential effect of COVID-19 exposure specifically for women.
To capture this effect, we include the interaction between being exposed to the COVID-19 pandemic and being an author that we identify as a woman.

In the end, the main equation that we estimate is the following: 
\begin{equation}
\begin{split}
    \mathit{Commits}_{acw} = & ~ \beta_1 \mathit{Woman}_a \times \mbox{\itshape Exposure to COVID-19}_{cw}  \\ 
    &+ \beta_2 \mbox{\itshape Exposure to COVID-19}_{cw} \\  
    &+ \gamma_a + \theta_w + \alpha Time \times \mathit{Woman}_{a} + \epsilon_{acw}    
\end{split}
\end{equation}
where:
\begin{itemize}
    \item  $\mathit{Commits}_{acw}$ is the number of commits of author $a$ identified as living in country $c$ during week $w$;
    \item $\mathit{Woman}_{a}$ is a binary variable equal to $1$ if author $a$ is identified as being a woman or $0$ otherwise;
    \item $\mbox{\itshape Exposure to COVID-19}_{cw}$ is a measure of exposure to the COVID-19 pandemic in country $c$ during week $w$. We consider two alternative measures: the number of COVID-19 attributed deaths and the value of the social distance stringency index (see Section~\ref{sec:covid-data} for the origin of these data);
    \item $\gamma_a$ are author fixed effects;
    \item $\theta_w$ are week fixed effects;
    \item $Time \times Woman_{a}$ are women specific trends; and
    \item $\epsilon_{acw}$ is the error term. 
\end{itemize}

\noindent
Author fixed effects capture all author-constant characteristics that could explain the results in part, including, but not limited to, those related to being a woman in a specific country.

Imagine, for example, that in developing countries, authors create fewer commits than in developed countries and that developing countries are affected by COVID-19 more than developed countries.
In this case, there would be a correlation between COVID-19 and the number of commits that is explained by different intrinsic characteristics of countries.
However, in the absence of fixed effects, our estimates of a naive regression would capture this correlation, and we could misinterpret a negative coefficient of COVID-19 on the number of commits as causal, which would not be the case.
The fixed effects that we include are not just at the country level but at the individual level, which allows us to get rid of any author-constant characteristics that could be correlated with COVID-19 (e.g., author-cohort, place of residence, or even gender).

More precisely, fixed effects subtract the mean of the fixed effects category.
For author fixed effects, the mean of the number of commits per week of each author is subtracted from each observation, thus making the authors more comparable between each other. 

Intuitively, a common way to interpret fixed effects is to imagine that the variation that is being used is within author, so that we compare the same author at different points in time. 

Week fixed effects capture all time-related changes that affect all authors in the same way, including global trends.
This could include for example the common effects of the WHO (World Health Organization) declaring a global pandemic for the whole planet, or a common decrease in economic activity across all countries.
Other non-pandemic specific changes, for example a lower number of commits during summer or the Christmas season would also be captured by week fixed effects. 

Finally, the women-specific trends capture time trends that are different for women.
Including women-specific trends is important in this case because during the time period we consider for our analysis, we observe both an increasing slope for the total number of commits done by women and the number of COVID-19 deaths.
Not controlling for a women-specific trend could result in misleading results, as we would correlate more COVID-19 deaths with more women contributions just because of a pre-existing trend. 

The estimate of interest is $\beta_1$ that captures the differential effect for women (with respect to comparable men) of being exposed to a certain level of COVID-19 in their country.

One assumption of classical ordinary least squares (OLS) regression is that all data points are independent of each other and are drawn from the same theoretical distribution.
In many settings, this is not the case.
Inhabitants of the same country, for example, share common institutions and characteristics that make them not independent of each other.
Similarly, individuals of the same gender or age also share common characteristics.
When this happens, some statistical methods allow relaxing the assumption that all observations are independent and identically distributed and determining some groups within which observations are not considered independent.
This is called clustering. 

Failing to recognize certain data points as not independent can lead to inaccurate standard errors and then to consider some coefficient estimates as statistically significant while in reality they are not.
Clustering standard errors is then crucial for statistical inference.
However, it is less clear at which level standard errors should be clustered. 
Recent econometric literature shows that for standard errors to be consistent\footnote{A \emph{consistent estimator} is an estimator that, when the number of observations goes to infinity, converges to the real value of the parameter for the entire population.} observations should be allowed to  be correlated between units that share the same treatment.

Following the recent literature on standard errors~\cite{abadie2023should} and DID estimators~\cite{bertrand2004much}, we cluster standard errors at the treatment level.
More precisely, we construct groups within which we allow correlation between observations.
Data points of the same country, gender, and period of treatment (before or after COVID-19) are in the same group.
Then, for each country, we have four different groups: one for women before the first week when any lockdown measure was implemented (or any COVID-19 related death had occurred); one for women after the week when the first lockdown measure was implemented in that country; and the same two groups for men. 

To answer \RQcause, we run the regressions corresponding to the empirical identification of equation (1) and cluster standard errors as described above using OLS with the statistical program STATA-17.
To answer \RQwho and analyze which categories of women are most affected, we run the same regression as before, but restricting the sample to different categories.
In the next section we describe the main result and detail further the analysis for different categories, showing obtained results.

\section{Results}
\label{sec:results}

In this section, we present the results of our empirical analysis.
Overall, the results show that higher exposure to lockdown measures due to COVID-19 reduced disproportionately the ability of women to contribute (compared to men).
We also show that, among women, contributors that tended to contribute more outside working hours and contributors outside academia and companies are the most affected.

\subsection{RQ1: differential effect of COVID-19 on women}

\begin{table}
  \caption{\textbf{Relative effect of COVID-19 on women's contributing activity, using two different measures of COVID-19 and two different measures of contributing activity}}
 
 \begin{center}
 \label{table:main}
 \begin{tabular}{lccccc}
 & (1) & (2) & & (3) & (4) \\
 & \multicolumn{2}{c}{\bfseries N.~of commits} & & \multicolumn{2}{c}{\bfseries N.~of active days}\\
 \cline{2-3} \cline{5-6} \\[1ex]
 \textbf{Woman $\quad \times$} & -0.00508** & & & -0.000589 &  \\
 \textbf{$\quad$ COVID-19 deaths} & (0.00245) & & & (0.000423) &  \\
 \textbf{$\quad$ (per \num{100000} inhab.)} & & &  &  & \\[2ex]
 \textbf{Woman $\quad \times$} & & -0.0882*** & & & -0.0136*** \\
 \textbf{$\quad$ social distance index} &  & (0.0240) & & & (0.00499) \\[2ex]
 \textbf{Author FE} & Yes & Yes & & Yes & Yes \\
 \textbf{Week FE} & Yes & Yes & & Yes & Yes \\
 \textbf{Gender trend} & Yes & Yes & & Yes & Yes \\[2ex]
 \textbf{Avg. dep.~var.} & 0.592 & 0.589 & & 0.150 & 0.149 \\
 \textbf{Avg. (women)} & 0.381 & 0.380 & & 0.106 & 0.106 \\
 \textbf{Avg. (men)} & 0.624 & 0.620 & & 0.156 & 0.156 \\[2ex]
 \textbf{Observations} & 1.310e+08 & 1.300e+08 & & 1.310e+08 & 1.300e+08 \\
 \textbf{N.~clusters} & 762 & 640 & & 762 & 640 \\ \hline
 \end{tabular}
 \end{center}
 
 Note: This table shows the relative effect of COVID-19 on women's contributing activity.
 Each column (numbered at the top for easier reference) corresponds to the results of a different regression. Columns 1 and 2 use number of commits as an outcome (dependent or Y variable), while columns 3 and 4 use the number of active days.
 Columns 1 and 3 use the number of COVID-19 deaths relative to the population as independent (or X) variable, while columns 2 and 4 use the social distance index.
 All columns include author fixed effects, week fixed effect, and worldwide gender-specific trends as indicated by the mention ``Yes'' in the corresponding line.
 The sample is all the authors that created at least one commit during 2019 that were localized in a country and that were assigned a gender.
 Woman is a binary variable indicating if the author is classified as woman.
 COVID-19 deaths (per \num{100000} inhabitants) is the number of deaths due to COVID-19 per \num{100000} inhabitants during a specific week in a specific country.
 The social distance index is a measure of the severity of the COVID-19 related social distance measures applied in each country each week. 
 ``Avg. dep.~var.'' is the average number of the dependent variable, i.e., number of commits per week for columns 1 and 2 or number of active days per week for columns 3 and 4, for our entire sample.
 To help interpret the results, we also provide the average number of the dependent variable for women authors and men authors in our sample.
 Standard errors are clustered at the gender-country-post treatment level.
 Stars represent different levels of statistical significance: ***~p$<$0.01; **~p$<$0.05; *~p$<$0.1.
 
 \end{table}

The main result of our analysis, the overall (relative) effect of COVID-19 on women's ability to contribute to public code, is shown in Table~\ref{table:main}.

We consider two different measures of exposure to COVID-19: (1) the number of COVID-19-related deaths per \num{100 000} inhabitants in a given week in a given country, and (2) the stringency of the lockdown measures implemented as a response to the COVID-19 pandemic in a given week in a given country (captured by the index of lockdown-style policies strictness described in Section~\ref{sec:covid-data}).
We also consider two different measures of the ability of an author to contribute: the number of commits authored in a week, and the number of active days in a week (i.e., the number of days in which the author authored at least one commit).

With both measures of COVID-19 exposure and both measures of ability to contribute, we obtain results leading to the same conclusion: \textbf{COVID-19 disproportionately affected women's ability to contribute to public code, compared to men}.

More precisely, in three out of four regressions in Table~\ref{table:main}, the coefficients of interest (the interaction between the \emph{Woman$_a$} binary variable and the two measures of COVID-19) are negative and statistically significant (at the 5\% level in column 1 and at the 1\% level in columns 2 and 4, which are the two columns reporting the results using the stringency index). For the last column (effect of the number of COVID-19 deaths on the number of active days), the coefficient is negative but not statistically significant.

Note that the magnitudes of our coefficients of interest do not represent the total effect of COVID-19 on women, but the \emph{difference} of the effect between men and women. Furthermore, because of author fixed effects, they do not allow comparing an average man with an average woman. Instead, we have to compare men and women authors that would have similar pre-pandemic contribution levels.

For column 1, the magnitudes of the estimates can be interpreted as follows. Consider a man and a woman that would have produced the same number of commits in a given week if no COVID-19 deaths had occurred in their country. Then, if instead one COVID-19 death per \num{100 000} inhabitants occurred, the woman would have produced 0.005 fewer commits than the man in this week. This relative decrease represents around 1.3\% of the average number of per week contributions of women (computed by dividing the previous value by the mean of the dependent variable for women only). Alternatively, one COVID-19 death per \num{100 000} inhabitants decreases the contribution of women (with respect to men) by around 2.1\% of the difference in the averages between men and women, widening the gap of contributions.

To give a better sense of scale, we can interpret the magnitudes evaluating the effect for a specific number of COVID-19 deaths. During the week with the most number of COVID-19 deaths in the USA, there were about 7 deaths per \num{100 000} inhabitants, which then represents a 9.3\% differential drop in women's contributions compared to the mean of women and around a 15\% increase in the average gap of contributions between men and women.

For column 2, magnitudes can be interpreted as follows. Changing the social distance index measure from no restrictions to the highest level of stringency
decreases the contribution of women (again, with respect to comparable men) by 0.088 commits per week.
This decrease represents around 23\% of the average per-week contribution of women and around 37\% of the average gap in the contributions between men and women. 

Only 13 out of 216 countries reached the highest level of social distance stringency index.
To give a better sense of the scale then, we can interpret the magnitudes for a change from no restrictions to a value of 0.75 (instead of 1) of the social distance stringency index, which was reached in 86\% of countries at some point in time.
The corresponding decrease in women's contributions (relative to comparable men) represents 17\% of the mean of women's contributions and 28\% of the gap between men and women.

For column 4, the magnitudes can be interpreted as follows. Changing the social distance index measure from no restrictions to the highest level of stringency decreases the number of active days of women (again, with respect to comparable men) by 0.0136 days per week.
This decrease represents around 13\% of the average number of active days for women contributors and around 27\% of the average gap in the number of active days between men and women.

Although the two measures that we use to assess the impact of COVID-19 and the two measures that we use to evaluate the authors' ability to contribute give similar results, in order to perform finer-grained analyses to answer RQ2, we focus on a main specification, which we will use on different subsamples of our main dataset (different categories of authors).

We choose to use social distancing and number of commits as our main specification. The social distance index is easier to interpret (as it varies from 0 to 1), and also better captures what is most likely to explain why COVID-19 had an impact on women's ability to contribute to public code: the lockdown measures and their consequences in terms of the time that women (and men) could spend on such type of activity. Furthermore, this is the specification for which we obtain the lower p-value and the larger coefficient of interest (relative to the mean of the dependent variable).
This makes more likely that we will have more statistical power (i.e., we will have a higher ability to capture existing effects) when reducing the number of observations to study the different subsamples.

\subsubsection{Robustness checks}
\label{sec:robustness}

Table~\ref{table:robustness} in the appendix shows the robustness checks that we have conducted. They are variations of our main specification (whose results are repeated in column 1), either by altering data pre-processing, or by changing the regression specification itself.

In columns 2 and 3, we assess the robustness of our geo-localization method by changing some of its parameters: column 2 includes more authors, by lifting the exclusions of 9 ccTLDs that we considered problematic for geo-localization and of emails from multinational companies; on the opposite, column 3 uses a much more restricted list of ccTLDs, by excluding any ccTLD appearing on the Wikipedia page ``Country code top-level domains with commercial licenses''\footnote{\url{https://en.wikipedia.org/wiki/Country_code_top-level_domains_with_commercial_licenses}, accessed November 2023} (in addition to our initial list of excluded ccTLDs), leading to a list of 55 exclusions, including large countries such as Iran and Italy.

In columns 4 to 6, we exclude some specific countries from our analysis to ensure that these countries are not exclusively driving our results. column 4 excludes the USA, which is both a large country with many contributors and one in which we could mostly identify contributors via university email domains (as the \texttt{.us} ccTLD is seldomly used); column 5 excludes countries which are outliers in terms of how many authors we identified relatively to their population (excluding both the top and bottom 5\% countries when sorting by number of authors localized in this country over the country population).
In particular, excluding the top 5\% is another way of making sure that we are not sensible to countries whose ccTLD would be abused for domain hacks (and where we would localize too many authors compared to the reality).
Finally, column 6 excludes countries for which we have too few women or men authors (less than 100), and which would therefore risk to lead to a biased comparison given the small number of observations. 

In column 7, we test an even more demanding variant of our main specification, where gender-specific trends can differ in each different country.
This alternative specification is likely to capture more gender-specific pre-pandemic variability due to varying politics across the world (e.g., while gender equality is slowly increasing in most countries, there are countries that experience setbacks).

The outcome for all these checks (shown in Table~\ref{table:robustness}) \textbf{confirm the robustness of our main specification: the result continues to hold in all cases, with the same level of statistical significance (at the 1\% level) and comparable coefficients}.

\subsection{RQ2: who are the women most affected by COVID-19}

To answer RQ2 we now re-run the regression for our main specification (social distancing index and number of weekly commits) on subsamples of our population corresponding to various categories of contributors, in order to understand which women were the most affected by the COVID-19 pandemic (relatively to comparable men).

\subsubsection{Distinguishing contributors by experience}

\begin{table}
\caption{\textbf{Relative effect of COVID-19 on women's number of weekly commits for different user categories, defined by year of first commit}}
\begin{center}
\label{table:cat-year-stringency}
\begin{tabular}{lcccc}
& (1) & (2) & (3) & (4) \\
\textbf{Year of first commit} & \textbf{all years} & \textbf{2019} & \textbf{2017--2018} & \textbf{before 2017} \\
\hline \\
& \multicolumn{4}{c}{\bfseries Number of commits}\\
\cline{2-5} &  &  &  &  \\
\textbf{Woman} $\quad \times$  & -0.0882*** & -0.0513** & -0.0357 & -0.206*** \\
$\quad$ \textbf{social distance index} & (0.0240) & (0.0249) & (0.0331) & (0.0489) \\[2ex]
\textbf{Author FE} & Yes & Yes & Yes & Yes \\
\textbf{Week FE} & Yes & Yes & Yes & Yes \\
\textbf{Gender trend} & Yes & Yes & Yes & Yes \\[2ex]
\textbf{Avg. dep.~var.} & 0.589 & 0.237 & 0.700 & 1.280 \\
\textbf{Avg. (women)} & 0.380 & 0.199 & 0.610 & 0.839 \\
\textbf{Avg. (men)} & 0.620 & 0.244 & 0.712 & 1.316 \\[2ex]
\textbf{Observations} & 1.300e+08 & 6.680e+07 & 3.450e+07 & 2.840e+07 \\
\textbf{N.~clusters} & 640 & 623 & 559 & 515 \\ \hline
\end{tabular}
\end{center}
Note: This table shows the relative effect of COVID-19 on women's contributing activity in different user categories depending on the year of first commit. Each column (numbered at the top for easier reference) corresponds to the results of a different regression.  The sample is all the authors that created at least one commit during 2019 that were localized in a country and that were assigned a gender. Woman is a binary variable indicating if the author is classified as woman. The social distance index is a measure of the severity of the COVID-19 related social distance measures applied in each country each week. All columns include author fixed effects, week fixed effect, and worldwide gender-specific trends, as indicated by a ``Yes'' in the corresponding column and line. Column 1 is our baseline specification for comparison. Column 2-4 restrict the sample to specific categories depending on the year of their first commit. Column 2 shows the results for authors that wrote their first commit in 2019; column 3, in 2017 or 2018; and column 4, in 2016 or before.
``Avg. dep.~var.'' is the average number of weekly commits for each subsample. To help interpret the results, we also provide the average number of weekly commits for women authors and men authors in each subsample. Standard errors are clustered at the gender-country-post treatment level. Stars represent different levels of statistical significance: ***~p$<$0.01; **~p$<$0.05; *~p$<$0.1.

\end{table}

We first categorize by contributor's experience.
For each author, we consider the year in which they created their first commit, and we create three categories out of this: started contributing in 2019 (the last year just prior to the pandemic), in 2017--2018, and before 2017.
We run the same regression for each category.
The results are presented in Table~\ref{table:cat-year-stringency}.
Column 1 just repeats the main result from Table~\ref{table:main}.
The first category (column 2) is authors who created their first commit to public code in 2019.
They represent $\approx\,$51\% of our dataset.
The second category (column 3) is authors who created their first commit in 2017--2018.
They account for $\approx\,$27\% of the dataset.
Our third category (column 4) is authors who created their first commit before 2017.
They represent $\approx\,$22\% of our dataset.
By just looking at the mean of the dependent variable (overall, and restricted to specific genders), we can observe that the more experienced a contributor is, the more commits they contribute every week on average.
(Note that our sample excludes authors that did not contribute any commit in 2019.)

The coefficients of interest are negative for all our categories of authors, and they are statistically significant for the first and last categories (authors with the least and the most experience).

Interpreting the magnitudes for the two categories with statistically significant coefficients, we can see that changing the social distance index from no restriction to the maximum level of restriction is responsible for a relative decrease of 0.206 weekly commits for the contributors with the most experience (column 4), compared to men that would have committed the same in a scenario without restrictions.
This decrease corresponds to 25\% of the average of weekly contributions for women in this category of more experienced users (compared to a 23\% decrease in the main result), and a 43\% increase in the gap between the average men and women levels of contribution (compared to a 37\% increase of the gap in the main result).

For contributors with the least experience (column 2), the relative decrease of weekly commits is of 0.0513, which represents 26\% of the average weekly contributions for women in this category, and a 114\% increase of the gap between the average men and women levels of contribution in this category.

Overall, we can see that COVID-19 affected contributors at different levels of experience, at a similar level of magnitude (relative to the average number of weekly contributions by women in each category), and \textbf{we cannot conclude that a specific category of women, by contributing experience, is driving our main result}.

\subsubsection{Distinguishing contributors by pre-pandemic levels of activity}

\begin{table}
  \caption{\textbf{Relative effect of COVID-19 on women's number of weekly commits for different user categories, defined by number of commits in 2019}}
  
  \begin{center}

  \label{table:cat-commits-stringency}
  \begin{tabular}{lcccc}
  & (1) & (2) & (3) & (4) \\
  \textbf{Number of commits in 2019} & \textbf{all values} & \textbf{$<$ 100} & \textbf{100--999} & \textbf{$>$ 999} \\
  \hline \\
  & \multicolumn{4}{c}{\bfseries Number of commits}\\
  \cline{2-5}
  &  &  &  &  \\
  \textbf{Woman $\quad \times$} & -0.0882*** & -0.0604*** & -0.511*** & -9.452*** \\
  \textbf{$\quad$ social distance index} & (0.0240) & (0.0166) & (0.139) & (2.953) \\[2ex]
  \textbf{Author FE} & Yes & Yes & Yes & Yes \\
  \textbf{Week FE} & Yes & Yes & Yes & Yes \\
  \textbf{Gender trend} & Yes & Yes & Yes & Yes \\ \\
  \textbf{Avg. dep.~var.} & 0.589 & 0.249 & 2.993 & 28.83 \\
  \textbf{Avg. (women)} & 0.380 & 0.202 & 2.357 & 20.08 \\
  \textbf{Avg. (men)} & 0.620 & 0.257 & 3.065 & 29.40 \\[2ex]
  \textbf{Observations} & 1.300e+08 & 1.180e+08 & 1.120e+07 & 468300 \\
  \textbf{N.~clusters} & 640 & 638 & 460 & 200 \\ \hline
  \end{tabular}
  \end{center}
  Note: This table shows the relative effect of COVID-19 on women's contributing activity in different user categories depending on the number of commits in 2019. Each column (numbered at the top for easier reference) corresponds to the results of a different regression. The sample is all the authors that created at least one commit during 2019 that were localized in a country and that were assigned a gender. Woman is a binary variable indicating if the author is classified as woman. The social distance index is a measure of the severity of the COVID-19 related social distance measures applied in each country each week. All columns include author fixed effects, week fixed effect, and worldwide gender-specific trends, as indicated by a ``Yes'' in the corresponding line and column. Column 1 is our baseline specification for comparison. Columns 2--4 restrict the sample to specific categories depending on the total number of commits of the author in 2019 (before the onset of the pandemic). Column 2 shows the results for authors that wrote less than 100 commits in 2019; column 3, between 100 and 999; and column 4, 1000 commits or more.
  ``Avg. dep.~var.'' is the average number of weekly commits for each subsample. To help interpret the results, we also provide the average number of weekly commits for women authors and men authors in each subsample.  Standard errors are clustered at the gender-country-post treatment level. Stars represent different levels of statistical significance: ***~p$<$0.01; **~p$<$0.05; *~p$<$0.1.
  
  \end{table}

We create a second categorization based on pre-pandemic contribution levels.
For each author, we consider the number of commits they created in 2019 (before the onset of the pandemic) and we create three categories out of this.
We run the same regression for each category.
The results are presented in Table~\ref{table:cat-commits-stringency}.
Column 1 just repeats the main result from Table~\ref{table:main}.
The first category (column 2) is authors who created fewer than 100 commits in 2019; they represent $\approx\,$91\% of the dataset.
The second category (column 3) is authors who created between 100 and 999 commits in 2019; $\approx\,$9\% of the dataset.
The third category (column 4) is authors who created at least 1000 commits in 2019; $\approx\,$0.4\% of the dataset.
We keep statistically significant results for all the categories.

Interpreting the magnitudes, we can see that, for the first category (authors who committed fewer than 100 commits in 2019), the relative decrease of women contributions in a week with the maximal level of social distancing is of 0.0604 weekly commits, which represents 30\% of the average weekly contribution for women in this category and a 110\% increase of the gap between the average men and women contributions.
For the second category (authors who committed between 100 and 999 commits in 2019), the relative decrease of women contributions in such a week is of 0.511 commits, which represents 22\% of the average weekly contribution for women in this category and a 72\% increase of the gap between the average men and women contributions.
For the third category (authors who committed at least 1000 commits in 2019), the relative decrease of women contributions in such a week is of 9.452 commits, which represents 47\% of the average weekly contribution for women in this category and a 101\% increase of the gap between the average men and women contributions.

Overall, we can see that COVID-19 affected contributors with different levels of pre-pandemic activity and, similarly to the previous categorization, \textbf{we cannot conclude that a specific category of women, by pre-pandemic levels of activity, is driving our main result}.

\subsubsection{Distinguishing professional and hobbyist contributors by contribution patterns}

\begin{table}
  \caption{\textbf{Relative effect of COVID-19 on women's number of weekly commits for different user categories, defined by percentage of commits during working hours}}
  \begin{center}

  \label{table:cat-hobby-stringency}
  \begin{tabular}{lccccc}
  \textbf{\% during} & (1) & (2) & (3) & (4) & (5) \\
  \textbf{working hours} & \textbf{all values} & \textbf{0\%} & \textbf{]0\%,50\%]} & \textbf{]50\%,100\%[} & \textbf{100\%} \\
  \hline \\
  & \multicolumn{5}{c}{\bfseries Number of commits}\\
  \cline{2-6} &  &  &  &  &  \\
  \textbf{Woman} $\quad \times$ & -0.0882*** & -0.0461*** & -0.170*** & -0.137*** & -0.00139 \\
  \textbf{~ social dist.} & (0.0240) & (0.0177) & (0.0507) & (0.0460) & (0.0247) \\
  \textbf{~ index} &  &  &  &  &  \\[2ex]
  \textbf{Author FE} & Yes & Yes & Yes & Yes & Yes \\
  \textbf{Week FE} & Yes & Yes & Yes & Yes & Yes \\
  \textbf{Gender trend} & Yes & Yes & Yes & Yes & Yes \\[2ex]
  \textbf{Avg. dep. var.} & 0.589 & 0.0913 & 0.787 & 1.071 & 0.130 \\
  \textbf{Avg. (women)} & 0.380 & 0.0698 & 0.515 & 0.720 & 0.117 \\
  \textbf{Avg. (men)} & 0.620 & 0.0950 & 0.825 & 1.120 & 0.132 \\[2ex]
  \textbf{Observations} & 1.300e+08 & 1.770e+07 & 3.640e+07 & 3.860e+07 & 3.710e+07 \\
  \textbf{N.~clusters} & 640 & 534 & 559 & 571 & 596 \\ \hline
  \end{tabular}
  \end{center}
  Note: This table shows the relative effect of COVID-19 on women's contributing activity in different user categories depending on the percentage of commits during working hours before 2020. Each column (numbered at the top for easier reference) corresponds to the results of a different regression. The sample is all the authors that created at least one commit during 2019 that were localized in a country and that were assigned a gender. Woman is a binary variable indicating if the author is classified as woman. The social distance index is a measure of the severity of the COVID-19 related social distance measures applied in each country each week. All columns include author fixed effects, week fixed effect, and worldwide gender-specific trends, as indicated by a ``Yes'' in the corresponding line and column. Column 1 is our baseline specification for comparison. Column 2--5 restrict the sample to specific categories, depending on the percentage of commits that each author wrote during working hours before 2020 (i.e., before the onset of the pandemic). Column 2 shows the results for authors that did not write any commits during working hours; column 3, for authors that wrote at least one commit during working hours but less than 50\% of their commits are during working hours. Column 4 shows the results for authors that wrote the majority of their commits but not all during working hours; and column 5, for authors that wrote all of their commits during working hours.
  ``Avg. dep. var.'' is the average number of weekly commits for each subsample. To help interpret the results, we also provide the average number of weekly commits for women authors and men authors in each subsample.  Standard errors are clustered at the gender-country-post treatment level. Stars represent different levels of statistical significance: ***~p$<$0.01; **~p$<$0.05; *~p$<$0.1.
  
  \end{table}

We create a third categorization to distinguish between professional and hobbyist contributors.
For each author, we consider the percentage of commits they created during working hours in our commit dataset before 2020 (i.e., between May 28th, 2018 and the end of 2019). We define working hours as 7.00AM--6.59PM from Monday to Friday in commits' local time.
We create four categories out of this.
We run the same regression for each category.
The results are presented in Table~\ref{table:cat-hobby-stringency}.
Column 1 just repeats the main result from Table~\ref{table:main}.
The first category (column 2) is authors who created all of their commits before 2020 outside working hours.
They represent $\approx\,$14\% of the dataset.
The second category (column 3) is authors who created more than half of their commits before 2020 outside working hours; $\approx\,$28\% of the dataset.
The third category (column 4) is authors who created more than half of their commits before 2020 during working hours; $\approx\,$30\% of the dataset.
The last category (column 5) is authors who created all of their commits before 2020 during working hours; $\approx\,$28\% of the dataset.

We obtain statistically significant results on a relative decrease of contributions of women due to COVID-19 for all categories except the last one (100\% of commits before 2020 during working hours).
Looking at the magnitudes allows us to understand a bit more which categories are most affected.
Indeed, just by comparing the middle two categories (columns 3 and 4), we can already observe that the coefficient of interest is higher in absolute value for the category of authors that commit more outside working hours even though the average number of weekly commits for this category is lower.

More specifically, looking at the magnitudes for the three categories for which we got statistically significant results, we observe that:
\begin{itemize}
    \item For the category of authors that create more than half of their commits during working hours (column 4), the relative decrease of weekly commits by women for a week with the highest levels of social distance measures is 0.137, which represents 19\% of the average number of weekly commits for women in this category, and an increase of 34\% of the gap between the average number of weekly commits of men and of women in this category.
    \item For the category of authors that create more than half of their commits outside working hours (column 3), the relative decrease of weekly commits by women for such a week is 0.170, which represents 33\% of the average number of weekly commits for women in this category, and 55\% of the gap between the average number of weekly commits of men and of women in this category.
    \item For the category of authors that create all their commits outside working hours (column 2), the relative decrease of weekly commits by women for such a week is 0.0461, which represents 66\% of the average number of weekly commits for women in this category, and 183\% of the gap between the average number of weekly commits of men and of women in this category.
\end{itemize}
A clear pattern emerges: \textbf{the more authors were contributing to public code outside of working hours before the pandemic, the more women were negatively impacted by the pandemic in their ability to contribute} (relatively to men in the same category and with the same pre-pandemic contribution levels).

\subsubsection{Distinguishing professional and hobbyist contributors by their use of a professional email address}

\begin{table}
  \caption{\textbf{Relative effect of COVID-19 on women's number of weekly commits for different user categories, defined by their use of an email address from a university or a company}}
 
 \begin{center}
 \label{table:university}
 \begin{tabular}{lcccc}
 & (1) & (2) & (3) & (4) \\
 & & & & \textbf{Non uni. \&} \\ 
 \textbf{Email domain} & \textbf{All} & \textbf{University} & \textbf{Company} & \textbf{non company} \\
 \hline \\
 & \multicolumn{4}{c}{\bfseries Number of commits}\\
 \cline{2-5} &  &  &  &  \\
 \textbf{Woman $\quad \times$} & -0.0882*** & -0.0492 & -0.0785 & -0.104*** \\
 \textbf{$\quad$ social dist. index} & (0.0240) & (0.0550) & (0.134) & (0.0255) \\[2ex]
 \textbf{Author FE} & Yes & Yes & Yes & Yes \\
 \textbf{Week FE} & Yes & Yes & Yes & Yes \\
 \textbf{Gender trend} & Yes & Yes & Yes & Yes \\[2ex]
 \textbf{Avg. dep.~var.} & 0.589 & 0.602 & 0.504 & 0.589 \\
 \textbf{Avg. (women)} & 0.380 & 0.425 & 0.421 & 0.366 \\
 \textbf{Avg. (men)} & 0.620 & 0.642 & 0.513 & 0.619 \\[2ex]
 \textbf{Observations} & 1.300e+08 & 2.350e+07 & 2.370e+06 & 1.040e+08 \\
 \textbf{N.~clusters} & 640 & 438 & 272 & 632 \\ \hline
 \end{tabular}
 \end{center}
 
 Note: This table shows the relative effect of COVID-19 on women's contributing activity in different user categories depending on whether they have a company or university email address. Each column (numbered at the top for easier reference) corresponds to the results of a different regression. The sample is all the authors that created at least one commit during 2019 that were localized in a country and that were assigned a gender. Woman is a binary variable indicating if the author is classified as woman. The social distance index is a measure of the severity of the COVID-19 related social distance measures applied in each country each week. All columns include author fixed effects, week fixed effect, and worldwide gender-specific trends, as indicated by a ``Yes'' in the corresponding line and column. Column 1 is our baseline specification for comparison. Column 2--4 restrict the sample to specific categories, depending on whether they were identified as belonging to a university, to a company or none of these based on their email address.  Column 2 shows the results for authors with a university email address, column 3 shows the results for authors with a company email address, while column 4 shows the results for the authors that were not identified as belonging to a university or a company.
 ``Avg. dep. var.'' is the average number of weekly commits for each subsample. To help interpret the results, we also provide the average number of weekly commits for women authors and men authors in each subsample.  Standard errors are clustered at the gender-country-post treatment level. Stars represent different levels of statistical significance: ***~p$<$0.01; **~p$<$0.05; *~p$<$0.1.

 \end{table}

We create a fourth categorization to distinguish between professional and hobbyist contributors, this time based on their use of a professional email address.
For each author, we consider whether we were able to match their email address to a university or a company, using data from Wikidata.
We create three categories, and we run the same regression for each category.
The results are presented in Table~\ref{table:university}.
Column 1 just repeats the main result from Table~\ref{table:main}.
The first category (column 2) is authors that use a university-based email address that we could recognize.
They represent $\approx\,$17\% of our dataset.
The second category (column 3) is authors that use a company-based email address that we could recognize.
They represent $\approx\,$1.7\% of our dataset.
The last category (column 4) is authors that use an email address that we could not recognize as being from a university or a company.

We only get statistically significant results for the last category (column 4).
The magnitude of the coefficient of interest is higher for this category compared to our main result, even though the average number of weekly commits in this category is slightly smaller.
From this, we conclude that our \textbf{main result is mostly driven by contributors outside universities and companies}, i.e., mostly driven by hobbyists. Note that by excluding university email addresses, we exclude both professional academics and students using their university email address. For most universities, it is impossible to distinguish between students and employees based on the email address alone.

Interpreting the magnitude for this last category, we observe that the relative decrease of women contributions for weeks with the highest levels of social distance measures is of 0.104 weekly commits, which represents 28\% of the average weekly commit number by women in this category and 41\% of the gap between the averages for men and women in this category.

\section{Discussion}
\label{sec:discussion}

In this section we discuss our empirical findings (from Section~\ref{sec:results}), providing plausible interpretations as of their origin, and evaluating threats to their validity.

\subsection{Interpretation of the findings}
\label{sec:interpretation}

Our main result (\RQcause, Table~\ref{table:main}) shows that \textbf{the COVID-19 pandemic induced (rather than being merely correlated with) an impairment in the ability of women to contribute to public code}, relatively to men.
The result holds for different ways of measuring the ability to contribute, and is statistically strong, both in terms of significance and robustness (see Section~\ref{sec:robustness}).

While the goal of this work was primarily to verify \emph{if} such a causal relationship existed and its magnitude, the breakdown by different groups of contributors that we conducted to answer \RQwho suggests interpretations as to where the impairment comes from.

On the one hand, our first and second categorizations (by experience and by pre-pandemic levels of activity) did not show a specific category of contributors being more affected than others. We still obtained statistically significant results showing women being disproportionately affected by the pandemic in virtually all of these groups. Looking into the magnitudes of coefficients, we observed some variations in the way certain categories were affected, but with no clear pattern suggesting that a specific category was more affected than others.

On the other hand, our third and fourth categorizations (by contribution patterns and by the use of a professional email address) revealed a clear difference in the way the pandemic affected hobbyist and professional contributors. Our two methods of distinguishing between these two types of contributors led to the same conclusion: \textbf{hobbyist contributors were more affected by the pandemic than professional contributors}.

In the case of the third categorization, the more women were contributing outside of working hours before the pandemic, the more affected they were, relatively to their previous contribution levels, and relatively to men in the same category.
We also did not obtain any statistically significant differential effect for authors that were contributing all their commits during working hours before the pandemic.

In the case of the fourth categorization, contributors outside academia and companies were significantly affected, but for contributors inside academia and companies, we did not obtain any statistically significant result showing a differential effect of the pandemic on women's versus men's contributions. This does not mean that COVID-19 did not affect these contributors, but rather that the effect was not significantly different for men and women.

Overall, and consistently with the literature in other domains, one possible explanation for these results is that women were disproportionately affected by additional burdens because of social distancing measures (e.g., taking care of children when schools are closed), and therefore had less spare time (in comparison with men with similar pre-pandemic contribution patterns) that they could allocate, in particular, to contributing to open source as a hobby.
On the other hand, women who contribute to public code as part of their work are not affected as much, because work during lockdown could (and did in many cases) continue.

In fact, a tradition of working from home has existed in software development since well before COVID-19.
It is plausible that this pre-existing work-from-home culture is partially responsible for the reduced, or lack of differential effect of the pandemic that we observe for professional contributors.
Developers that were \emph{already} remote working have been impacted nonetheless by the pandemic~\cite{Bao2022, Butler2021, Ford2021, McDermott2021} (as discussed in detail in Section~\ref{sec:related}).

For the case of our negative result for contributors with university email addresses, we can speculate different explanations for students and university employees that are consistent with our results.
For university employees, the explanation is most likely the same as for contributors from companies: they continued to work during the lockdown, and therefore their ability to contribute to public code was not affected by the pandemic (relatively to men).
Regarding students, women students are less likely to be disproportionately affected (relatively to their men counterparts) by social distancing measures because most of them do not yet have children or elderly parents in their care.
For some students, the pandemic could even have increased the spare time that they could allocate to contributing to public code, but there is no reason that this effect would be different for men and women students.

\subsection{Threats to validity}
\label{sec:threats}

In the remainder of this section we discuss threats to the validity of the reported findings and the mitigations put in place to counter them.
We follow the terminology and threats classification by Runeson et al.~\cite{runeson2009guidelines}.

\subsubsection{Construct validity}

\paragraph{Analysis scale.}
We have conducted a large-scale analysis that encompasses, after data cleanup and identity merging, a set of \dataAuthorsPostMergeApprox authors.
At this scale, manual approaches (e.g., interviews) are not viable to obtain information about the gender or geographic location of the entire population.
Joining identities from the initial dataset with external platforms (e.g., social media websites) was also not feasible due to the lack of a viable join key.
We have hence resorted to fully automated methods for detecting gender and geographic location, in part reusing tools and techniques from recent works in the literature (for gender detection at this scale) and in part exploiting signals that we expect to be reliable (e.g., university email domains).
We therefore do not consider that the chosen techniques for feature detection pose a significant risk to the validity of the obtained results.
Nonetheless, we critically review and discuss below specific aspects of our choices for gender detection and geolocation.

\paragraph{Gender detection.}
We largely share the threats to the validity of gender detection of Rossi and Zacchiroli~\cite{icse-seis-2022-gender}, which we briefly recall here.
Several automated tools and services for name-based gender detection exist.
Among them, we have chosen to build upon \PKGGG for two main reasons: (1) it is open source, which not only enables better replicability and verifiability, but also allows scaling to the number of authors in our dataset without incurring per-query costs; (2) according to the benchmark conducted by Santamaria et al.~\cite{santamaria2018genderapi}, \PKGGG performs comparatively well with diverse datasets containing authors from all over the world, like ours.
Given \PKGGG works on \emph{first} names, our methodology adds on top of it two levels of majority assessment for determining the gender of an author: one among the tokens of each of their identities, another among the multiple identities of the same author.
One can easily construct artificial cases in which legitimate family names skew the gender of an author in the wrong direction, but this does not appear to happen in samples of the dataset that we have manually verified.

We were unable to determine the gender of significant parts of the dataset, but this is common for such large-scale experiments.
Also, the remaining sample that we have analyzed is still very significant in size, especially w.r.t.~most of the studies in the field.

\paragraph{Geolocation.}
We have used two different techniques for geolocation, both based on author emails: country inference based on country code top-level domains (ccTLDs) and inference based on university domains.

Both techniques can be gamed by authors, as emails in Git commits are not verified, but there is little incentive to do that just to be assigned to a different country.
Domain hacks (e.g., \texttt{.io} ccTLD for technology-related domains or \texttt{.me} for personal websites) will result in misclassifications,
which we have accounted for by excluding ccTLDs that are commonly used outside their respective countries.
We also performed a robustness check, where we exclude an even larger sample of ccTLDs that could be misused.
On the other hand, for historical reasons people in the US tend to use much less their own ccTLD domain (\texttt{.us}) than people located elsewhere in the world; such population in our experiment will be geolocated in the US mainly via the university domain technique, possibly resulting in their under-representation. For this reason, we also performed a robustness check where authors from the US were excluded.

We have verified the reliability of geolocation in two ways.
First, we have computed the agreement between the two geolocation techniques and found it to be very high: within the population of authors having contributed at least one commit in 2019 for which we could detect the gender and that were geolocated with \emph{both} techniques, only 40 unique authors were assigned to a different country. We have excluded these authors from the analysis.

Second, we have compared our geolocation results to the self-declared location of authors on GitHub, using as ground truth the latest available GHTorrent~\cite{GHTorrent} data dump, obtaining high accuracy (see Section~\ref{sec:dataset-code} for details).
These results instill confidence in our geo-localization technique, in spite of the amount of unclassifiable entries.
Notably, the majority of unclassified email addresses belong to familiar commercial domains such as Gmail, Hotmail, Yahoo, and anonymized GitHub addresses, totaling 66\% (\num{441 938} out of \num{671 766}) of all distinct email domains in the validation dataset.

As with all classification methods that introduce data filtering, there is a risk that filtering introduces a bias with respect to the dimensions of interest for the study that is conducted.
This seems very unlikely in this case: while the filtering can introduce a gender and/or a geographic bias to the data, we are interested in the temporal evolution of these data and the biases are unlikely to change through time.

\subsubsection{External validity}

\paragraph{Public code coverage.}
We started from a very large, but necessarily incomplete, sample of the entire body of public code and its version control history.
No archive of public code can be complete, and we inherit the incompleteness of the archive we used as starting point: Software Heritage.
To the best of our knowledge, Software Heritage is the largest publicly accessible archive of public code, which we believe validates our choice of it as the initial dataset.

Nonetheless, we acknowledge that this implies our analysis is affected by the archival coverage, or lack thereof, that Software Heritage has of public software development platforms.
In terms of archived platforms, Software Heritage archives the most popular development forges (e.g., GitHub and GitLab.com), as well as a number of less popular ones,\footnote{See \url{https://archive.softwareheritage.org} for coverage details, accessed December 2023.} so we consider the risk of missing platforms entirely to be limited.
But Software Heritage might also be missing relevant development activities (most importantly for our methodology: commits) due to various reasons.
We have discussed this with the archive operators and learned that there is indeed an archival lag w.r.t.~GitHub and we have taken it into account.
Specifically, we have verified that archive operators consider the data dump we
started from (March 18th, 2024) to be complete up to May 30th, 2022, and
restricted our analysis to that date: which was more than enough for our needs,
covering more than 2 years from the beginning of the pandemic (early 2020).

\paragraph{Gender identification.}
Name-based gender identification is not a perfect method and its application could result in some form of bias because of technical and social related issues.
In the data processing described in Section~\ref{sec:dataset-code}, we were able to associate a gender only to a part of the authors, leaving several in the \emph{unknown} class.
Our observations about that \emph{unknown} population support the idea that they mostly consist of a specific, well-characterized type of contributors, rather than a group of contributors that we randomly could not assign to a gender.
Focusing on the May 2018--May 2022 period, while the unknown active authors account for \dataPercUAuthors of the total, the commits they performed in that time span amount to \dataPercCommitsByUAuthors of the total; similarly, the average number of commits performed by unknown authors is \dataAvgCommitsUAuthors, which is significantly lower than the \dataAvgCommitsFAuthors of women and \dataAvgCommitsMAuthors of men.
One possible explanation for this difference is that these unknown are largely constituted by authors that did not configure a Git client with their name, e.g., because they are committing directly from a GitHub account.
In fact, when looking at the authors that we are able to geolocate, there are fewer authors for which we could not detect the gender: 38\% for authors localized with their email ccTLD and 40\% for authors localized with their university email domain.

Due to systemic online discrimination and harassment, women can take steps to hide their gender in public interactions, including code contributions.
They can for instance use a pseudonym or adopt a male-looking name.
This will result in under-counting women authors and their contributions (if they adopt a pseudonym that will be detected as ``unknown'' by \PKGGG) and possibly also in over-counting men authors and their contributions (if they adopt a male name or a pseudonym that will be detected as ``male'').
No mitigation is possible for this in the realm of name-based gender detection.
We only observe that we consider unlikely that this phenomenon has changed significantly across the pandemic and that the worst case (in terms of validity risk), women adopting male names, is less likely than other cases.

\paragraph{Non geo-localizable developers.}
The analysis we conducted is restricted to developers that we could geolocate with either of the email-based methods we have used.
It is possible that the resulting population subset is not representative of global tendencies, and also that country-specific patterns have been excluded from our analysis due to non-geolocated developers in those countries.
This risk cannot be avoided entirely, hence we make no claims about the generality of our claims outside the analyzed population sample.
Nonetheless, we conducted and reported about robustness checks (see Section~\ref{sec:robustness}) to verify this, not finding any issue.

\subsubsection{Reliability}
\label{sec:threats-reliability}

Reliability threats concern the extent to which our analyses and results depend on the team obtaining them.
As it is common practice, we mitigate this risk by providing a replication package for the work presented in this paper~\cite{this-replication-package}.

Note that, due to the presence of personal information in the dataset (names and emails) we are unable to provide a \emph{complete} replication package, in particular we have excluded such information from the replication package.
As the next best alternative, we include in the replication package the list of commit identifiers (in the form of SWHIDs~\cite{swhipres2018}) that constitute our starting public code dataset in Figure~\ref{fig:dataset-swh}.
This enables independent crawling of the same information, as well as cross-referencing the relevant commits with other datasets.

From there on, all other results are verifiable and reproducible using the replication package.

\section{Conclusion and future work}
\label{sec:conclusion}
\label{sec:future}

To act in favor of more inclusion of women in free/open source software (FOSS) development, it is crucial to understand the causes of their current underrepresentation.
In this work, we have investigated one such (recent and worldwide) cause, which had not been studied before in the context of FOSS development: the arrival of the COVID-19 pandemic.

By using a large dataset of contributors and contributions to public code, obtained from the Software Heritage archive, and difference in differences (DID), we have shown that there exists a causal relationship between the pandemic and a disproportionate reduction in women's ability to contribute to public code,  relative to men.
Furthermore, we have looked into which groups of women contributors have been most impacted by the pandemic, finding that it is specifically among hobbyist contributors that women have been most disproportionately affected.

These results are consistent with those obtained for contexts in society other than computing and open source, showing how systemic household disparities have been heightened by the pandemic.
We argue that open source communities who are committed to reducing gender inequalities need to take into account such systemic disparities and their exacerbation by external shocks, like the COVID-19 pandemic, when designing, implementing, and monitoring policies aimed at increasing diversity and inclusion.

\paragraph{Future work.}

Several aspects of this phenomenon remain to be investigated as future work.

Several years have now passed since the beginning of the COVID-19 pandemic.
Stringency measures to counter it have been weakened around the world, for better or worse.
It is now possible to verify whether doing so has enacted a ``bounce back'' in the ability of women to contribute to public code, relative to men.
While similar large-scale analyses should be possible, the methodology we have used in this work is not directly applicable to this new question, as it directly relates weeks with COVID-19 stringency measures to the contribution activity of authors in the same week. Thus, this method is not suited to look into longer-term effects of the pandemic on women's ability to contribute to public code.

In terms of policy factors, it would be interesting to know if some open source communities were better able to face the COVID-19 impact on women participation than others.
To verify that, the subgroup analysis that we conducted in this paper can be refined to partition contributors by the usual project characteristics (size, longevity, technological stack, etc.) as well as by relevant factors for the problem scope, e.g., the adoption of code of conducts (CoC) or other DEI (Diversity, Equity, Inclusion) related actions.
This finer-grained quantitative analysis could be complemented by more qualitative studies, e.g., interviews or surveys with maintainers and contributors of projects that have been particularly successful in maintaining or increasing the level of women participation despite the pandemic, to understand what practices have been effective.

Finally, we would like to look into the nature of the work done by contributors in our dataset.
Were women contributors working on specific tasks (e.g., development, testing, documenting, quality assurance, etc.) more impacted than women working on others?
And, if so, why is that the case?
Insights obtained on these questions can help make FOSS communities more resilient to future major external shocks, like COVID-19 has been.

\section*{Conflicts of Interest}

The authors declare that they have no conflict of interest.

\clearpage
\appendix
\section{Robustness checks results}
\label{sec:robustness-data}

Table~\ref{table:robustness} shows the outcomes of the robustness checks we have conducted on our main result.
(See Section~\ref{sec:robustness} for a discussion of these checks.)

\begin{landscape}
  \begin{table}
    \caption{\textbf{Robustness checks on the main specification: Relative effect of COVID-19 on women's number of weekly commits}}
   
   \begin{center}
   \label{table:robustness}
   \begin{tabular}{lcccccccc}
   & (1) & (2) & (3) & (4) & (5) & (6) & (7) \\
   & & & \textbf{no commerc.} & \textbf{excl.} & \textbf{excl.} & \textbf{excl.} & \textbf{country-gender} \\
   \textbf{Variant} & \textbf{main} & \textbf{no exclusion} & \textbf{ccTLD} & \textbf{USA} & \textbf{low repr.} & \textbf{few obs.} & \textbf{trend}
   \\ \hline
   \\
   & \multicolumn{7}{c}{\bfseries Number of commits}\\
   \cline{2-8}
   &  &  &  &  &  &  &  \\
   \textbf{Woman} $\quad \times$ & -0.0882*** & -0.0898*** & -0.0909*** & -0.0845*** & -0.0877*** & -0.0961*** & -0.0828*** \\
   \textbf{$\quad$ social distance index} & (0.0240) & (0.0234) & (0.0267) & (0.0252) & (0.0240) & (0.0236) & (0.0209) \\[2ex]
   \textbf{Author FE} & Yes & Yes & Yes & Yes & Yes & Yes & Yes \\
   \textbf{Week FE} & Yes & Yes & Yes & Yes & Yes & Yes & Yes \\
   \textbf{Gender trend} & Yes & Yes & Yes & Yes & Yes & Yes & No \\
   \textbf{Country-gender trend} & No & No & No & No & No & No & Yes \\[2ex]
   \textbf{Avg. dep.~var.} & 0.589 & 0.586 & 0.572 & 0.592 & 0.588 & 0.592 & 0.589 \\
   \textbf{Avg. (women)} & 0.380 & 0.379 & 0.365 & 0.374 & 0.380 & 0.380 & 0.380 \\
   \textbf{Avg. (men)} & 0.620 & 0.616 & 0.602 & 0.622 & 0.619 & 0.624 & 0.620 \\[2ex]
   \textbf{Observations} & 1.300e+08 & 1.320e+08 & 1.070e+08 & 1.160e+08 & 1.290e+08 & 1.260e+08 & 1.300e+08 \\
   \textbf{N.~clusters} & 640 & 642 & 518 & 636 & 604 & 220 & 640 \\ \hline
   \end{tabular}
   \end{center}
   
   Note: This table shows different robustness checks of the main specification on the relative effect of COVID-19 on women's contributing activity. Each column (numbered at the top for easier reference) corresponds to the results of a different regression that captures a different robustness check.The sample is all the authors that created at least one commit during 2019 that were localized in a country and that were assigned a gender. Woman is a binary variable indicating if the author is classified as woman. The social distance index is a measure of the severity of the COVID-19 related social distance measures applied in each country each week. All columns include author and week fixed effect, as indicated by a ``Yes'' in the corresponding line and column. Column 1 is our baseline specification for comparison. Columns 2--7 apply different variations to the specification for robustness checks. Columns 1--6 include worldwide gender-specific trends.
   Columns 2 and 3 change the list of ccTLDs that are excluded from our geo-localization method: column 2 does not exclude any ccTLD nor any multi-localized company, while column 3 excludes additional ccTLDs (on top of the initial list) using the list from the \href{https://en.wikipedia.org/wiki/Country_code_top-level_domains_with_commercial_licenses}{``Country code top-level domains with commercial license'' Wikipedia page} (accessed December 2023).
   Column 4--6 restrict our sample by excluding some countries: column 4 exclude the USA; column 5 excludes countries that are insufficiently representative of their population. In particular, we sort countries by the ratio of authors appearing  in our dataset of gendered and localized authors over the total population of the country, and we eliminate the top and the bottom 5\%. Column 6 excludes countries with too few observations. In particular, we eliminate countries with less than 100 women (or men) authors.
   Column 7 shows the results when using country-specific gender-specific trends.
   Standard errors are clustered at the gender-country-post treatment level. Stars represent different levels of statistical significance: ***~p$<$0.01; **~p$<$0.05; *~p$<$0.1.
   
   \end{table}
\end{landscape}

\end{document}